\documentstyle[12pt,epsfig]{article}

\def\WW{\mathrm{W}\mathrm{W}}
\def\qqqq{\mathrm{q}\overline{\mathrm{q}}\mathrm{q}\overline{\mathrm{q}}}
\def\qq{\mathrm{q}\overline{\mathrm{q}}}
\def\pp{\mathrm{p}\overline{\mathrm{p}}}
\def\qqln{\mathrm{q}\overline{\mathrm{q}}\ell\nu}
\def\lnln{\ell\nu\ell\nu}
\def\ipb{pb$^{-1}$}

\def\kag{\kappa_{\gamma}}
\def\goz{g^Z_1}
\def\lag{\lambda_{\gamma}}
\def\gw{\Gamma_W}
\def\Vcs{V_{cs}}

\def\beq{\begin{equation}}
\def\eeq#1{\label{#1}\end{equation}}
\def\eeqn{\end{equation}}

\def\beqa{\begin{eqnarray}}
\def\eeqa#1{\label{#1}\end{eqnarray}}
\def\eeqan{\end{eqnarray}}

\let\bar=\overbar

\def\Dslash{\not{\hbox{\kern-4pt $D$}}}
\def\dslash{\not{\hbox{\kern-2pt $\del$}}}

\def\ee{e^+e^-}

\def\mz{m_Z}

\def\mw{m_W}
\def\mt{m_t}

\def\msb{{\bar{\ssstyle M \kern -1pt S}}}

\def\Title#1{\begin{center} {\Large {\bf #1} } \end{center}}

\begin{document}

\begin{flushright}
BHAM-HEP/99-08
\end{flushright}

\Title{Precise Electroweak Results from LEP2}

\bigskip\bigskip


\begin{raggedright}  

{\it D.G. Charlton\index{Charlton, D.G.}\\
Royal Society University Research Fellow\\
School of Physics and Astronomy\\
The University of Birmingham\\
Birmingham B15 2TT, United Kingdom}
\bigskip\bigskip
\end{raggedright}

\section{Introduction}

The high luminosity delivered by LEP after the doubling of the 
$\ee$ collision energy means that LEP2 is now providing
substantial samples of W bosons with which to make complementary tests
of the Standard Model to those of LEP1.
This, together with collision energies in excess of 200~GeV, is
ensuring that the three central physics goals of the LEP2 programme
are properly explored: the precise measurement of the W mass; 
the measurement of vector-boson self-interactions;
and the search for new particles.
This review discusses the first two of these objectives: the
third is addressed elsewhere in these
proceedings\,\cite{vanina}.

\begin{table}[b]
\begin{center}
\begin{tabular}{l|cccccccc}  
Centre-of-mass energy (GeV) & 161 & 172 & 183 & 189 & 192 & 196 & 200
 & 202 \\
\hline
Integ. luminosity (\ipb) & 10 & 10 & 55 & 180 & 30 & 80 & 80 & 40 \\
Collection year & \multicolumn{2}{c}{1996} & 1997 & 1998 & 
\multicolumn{4}{c}{---~~~1999~~~---} \\
\end{tabular}
\caption{Accumulated data samples at LEP2. Integrated luminosity
totals are quoted per experiment, and are approximate.}
\label{tab:lumi}
\end{center}
\end{table}

\begin{figure}[thb]
\begin{center}
\epsfig{file=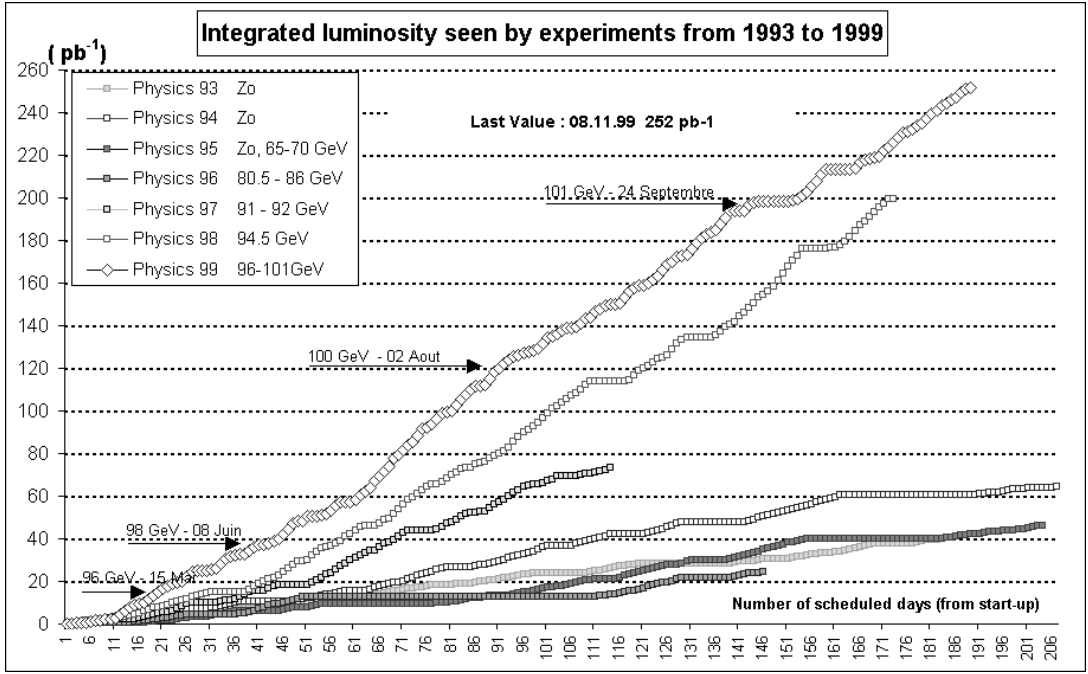,height=3.5in}
\caption{Integrated luminosity delivered per experiment, in the later years of
LEP1 and so far during LEP2.}
\label{fig:lumi}
\end{center}
\end{figure}

After a hesitant start of LEP above the W pair production threshold in
1996, subsequent years have seen increasingly large data samples
accumulated by the experiments (Table~\ref{tab:lumi}).
A total of around 480~\ipb\ has now been recorded by each experiment
at LEP2.
This integrated exposure can be expected to pass 650~\ipb\ by the
time data-taking is completed in the second half of 2000.
During 1999, the collision energy has reached, 
and surpassed\footnote{At the time
of the conference, the LEP collision energy had just reached
200~GeV. The figures quoted in the text refer to the achievements at
the time of writing (November 1999).}, its design value 
of $\sqrt{s}=200$~GeV.
As illustrated in Figure~\ref{fig:lumi}, the machine
performance has been better than ever in 1999, in spite of the
increased load on the RF system imposed by the higher energies.

This report describes various Standard Model tests made with
the first parts of the LEP2 data. 
Because analyses are in varying stages of completion essentially all
results quoted are preliminary.
In addition to the measurements made
with W pairs, studies of fermion-pair production, QED tests, and Z
pair production are reviewed.

\section{Fermion-Pair Production}

\begin{figure}[thb]
\begin{center}
\raisebox{2.4in}{a)}
\epsfig{file=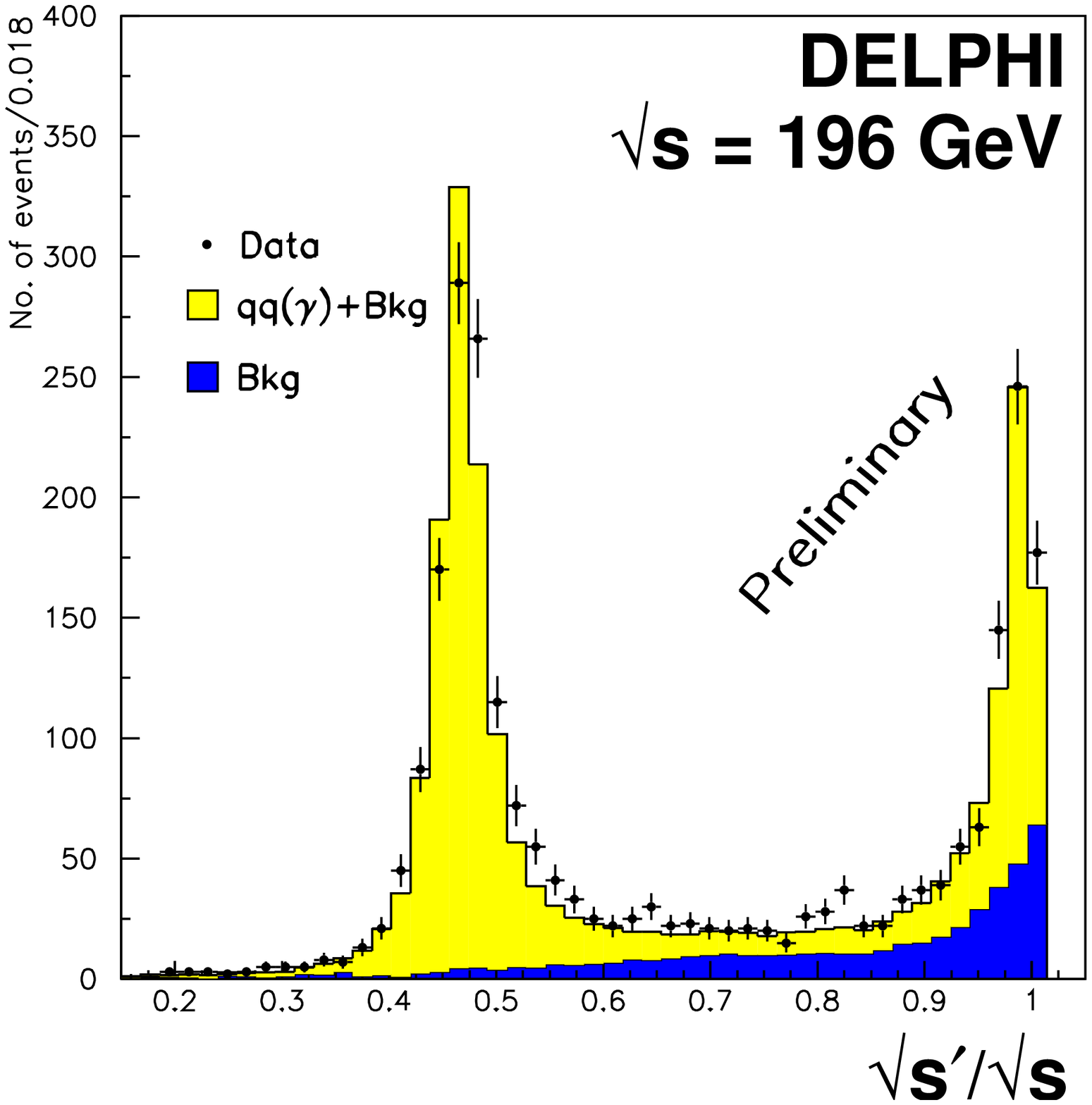,height=2.5in}
\raisebox{2.4in}{b)}
\epsfig{file=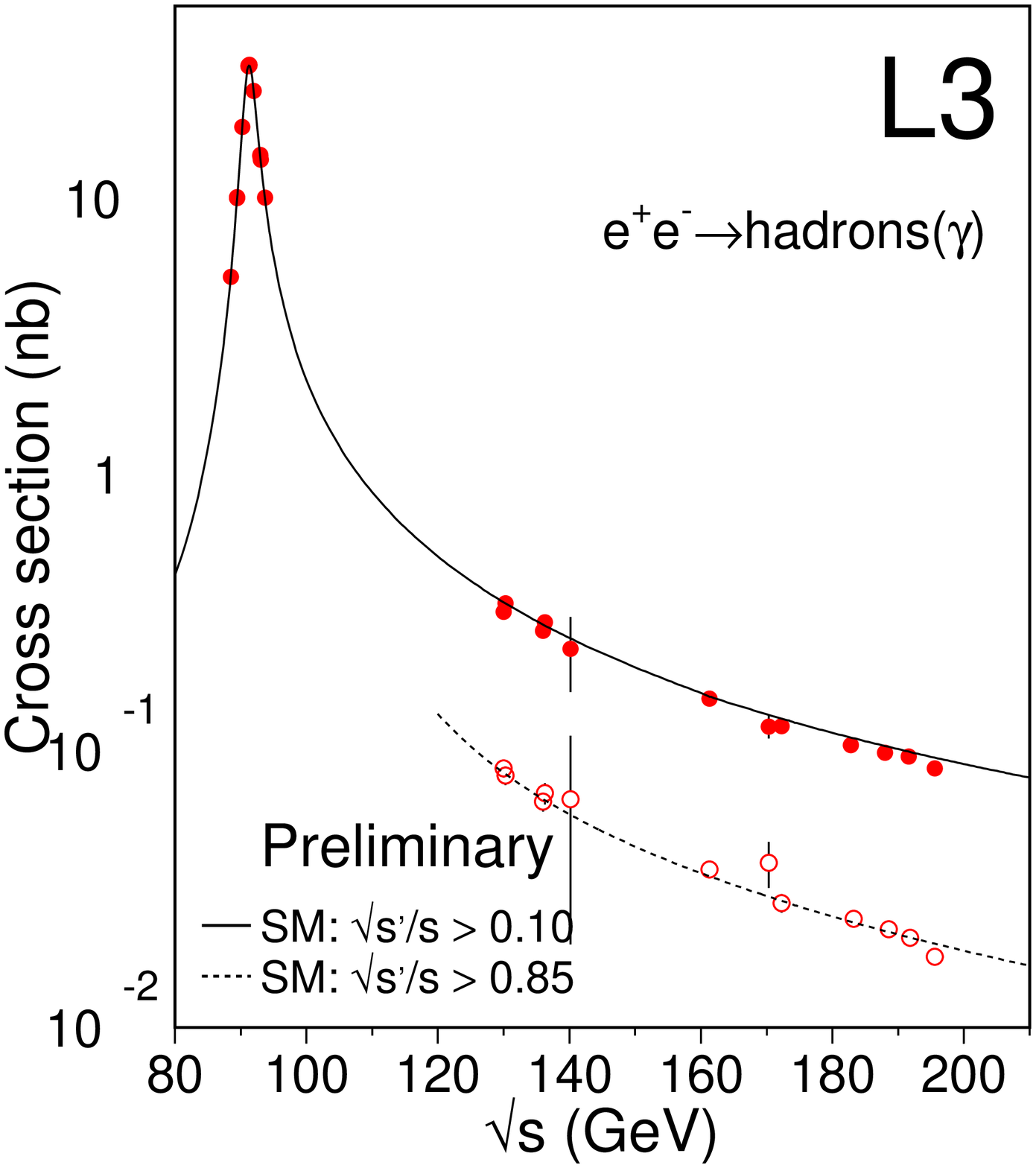,height=2.5in,width=2.6in}
\caption{a) Typical reconstructed $\sqrt{s'}/\sqrt{s}$
distribution\,\protect\cite{delphitwofermion}; 
b) Measured cross-sections for hadronic events from 
$\qq$ production\,\protect\cite{l3twofermion}.}
\label{fig:ffxs}
\end{center}
\begin{center}
\raisebox{2.4in}{a)}
\epsfig{file=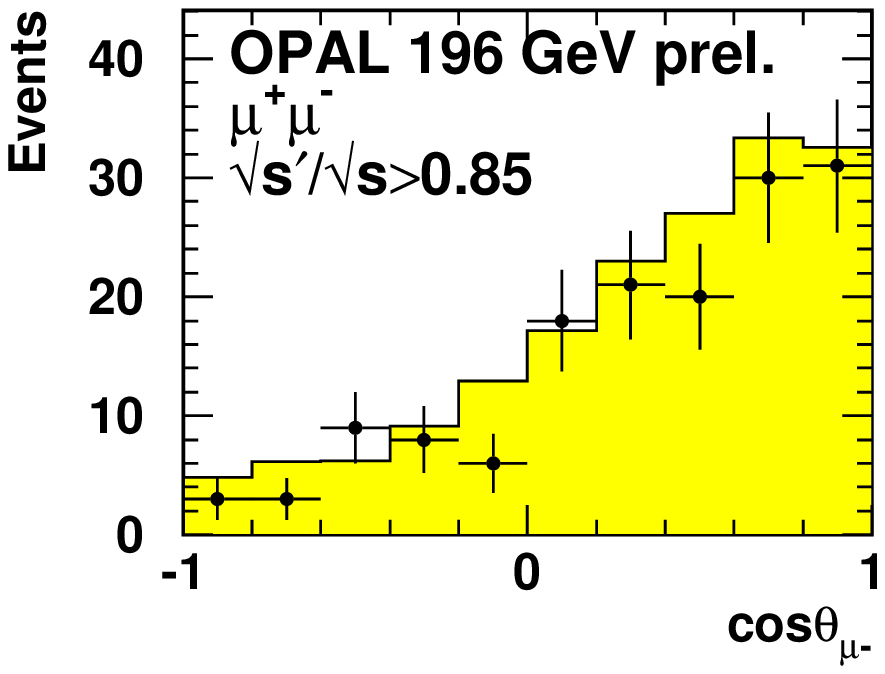,height=2.5in}
\raisebox{2.4in}{b)}
\epsfig{file=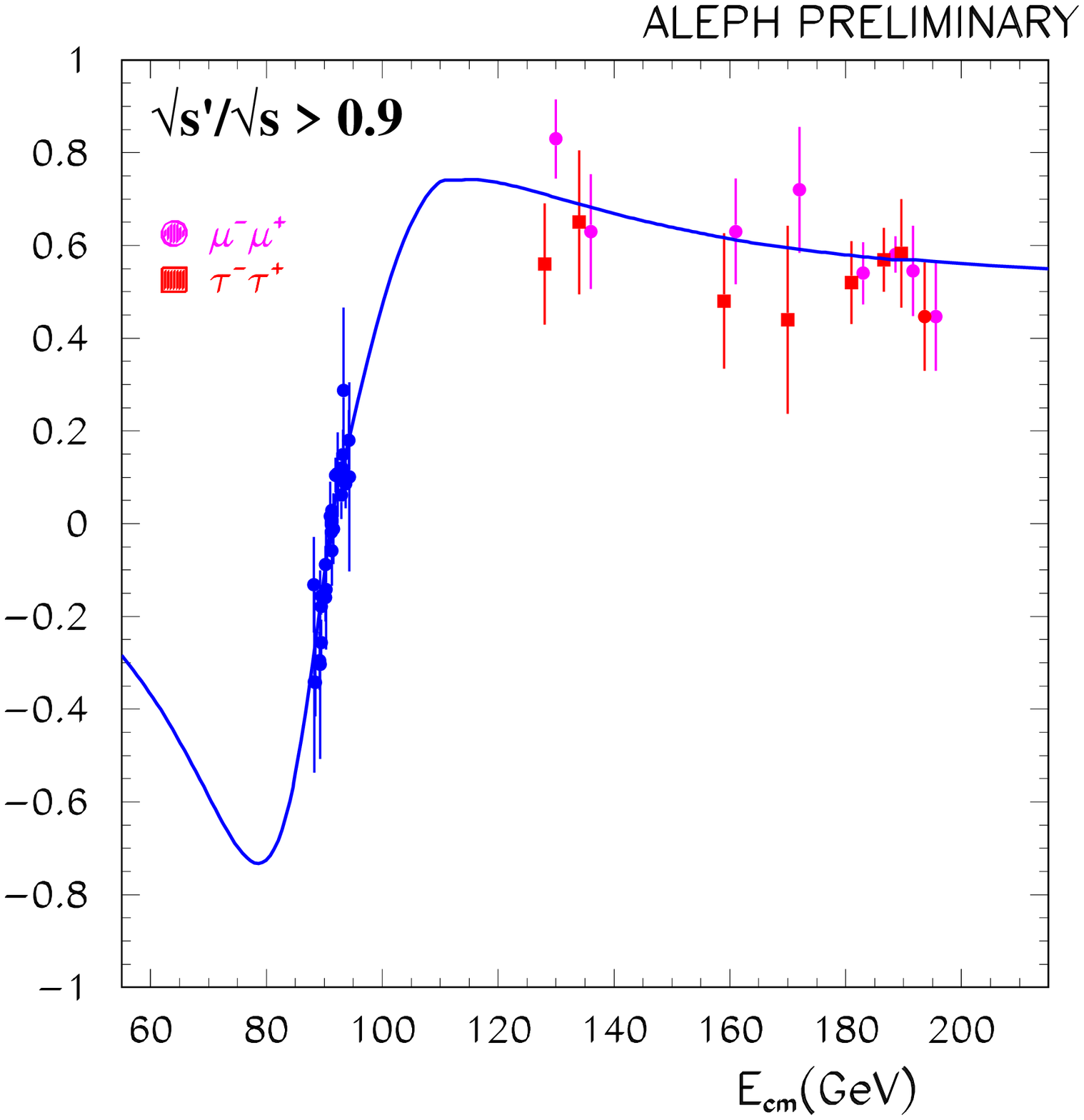,height=2.5in}
\caption{a) Typical reconstructed $\cos\theta_{\mu^-}$ 
distribution\,\protect\cite{opaltwofermion}; 
b) Measured asymmetries for muon and tau pair
production\,\protect\cite{alephtwofermion}.}
\label{fig:ffasy}
\end{center}
\end{figure}

Although two to three orders of magnitudes less than at LEP1
(Figure~\ref{fig:ffxs}b), the cross-section for fermion-pair
production at LEP2 is still high compared to many other processes.
The presence of the Z resonance at lower centre-of-mass energies
strongly affects the characteristics of events at higher energies,
because initial-state photon radiation leads to so-called ``radiative
return'' events where the fermion-pair system has an invariant mass
($\sqrt{s'}$) close to the Z. As a result two typical populations of
fermion-pair
events are observed, as illustrated in Figure~\ref{fig:ffxs}a: the
radiative return events with $\sqrt{s'} \simeq \mz$, and non-radiative
events with $\sqrt{s'} \simeq \sqrt{s}$.
The latter events are of more interest, as they probe the full
centre-of-mass energy scale.

\begin{figure}[thb]
\begin{center}
\raisebox{2.5in}{a)}
\epsfig{file=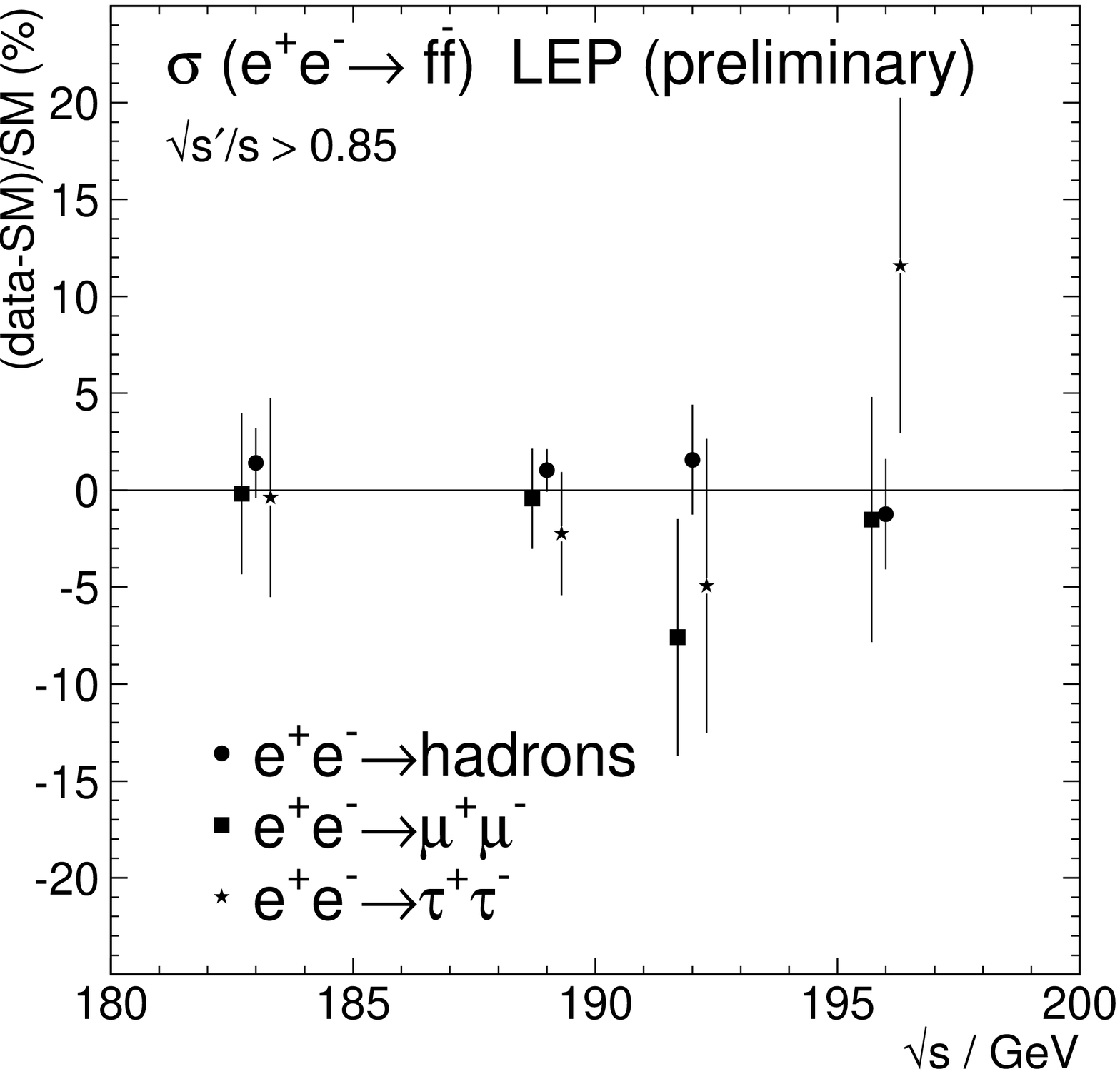,height=2.6in}
\raisebox{2.5in}{b)}
\epsfig{file=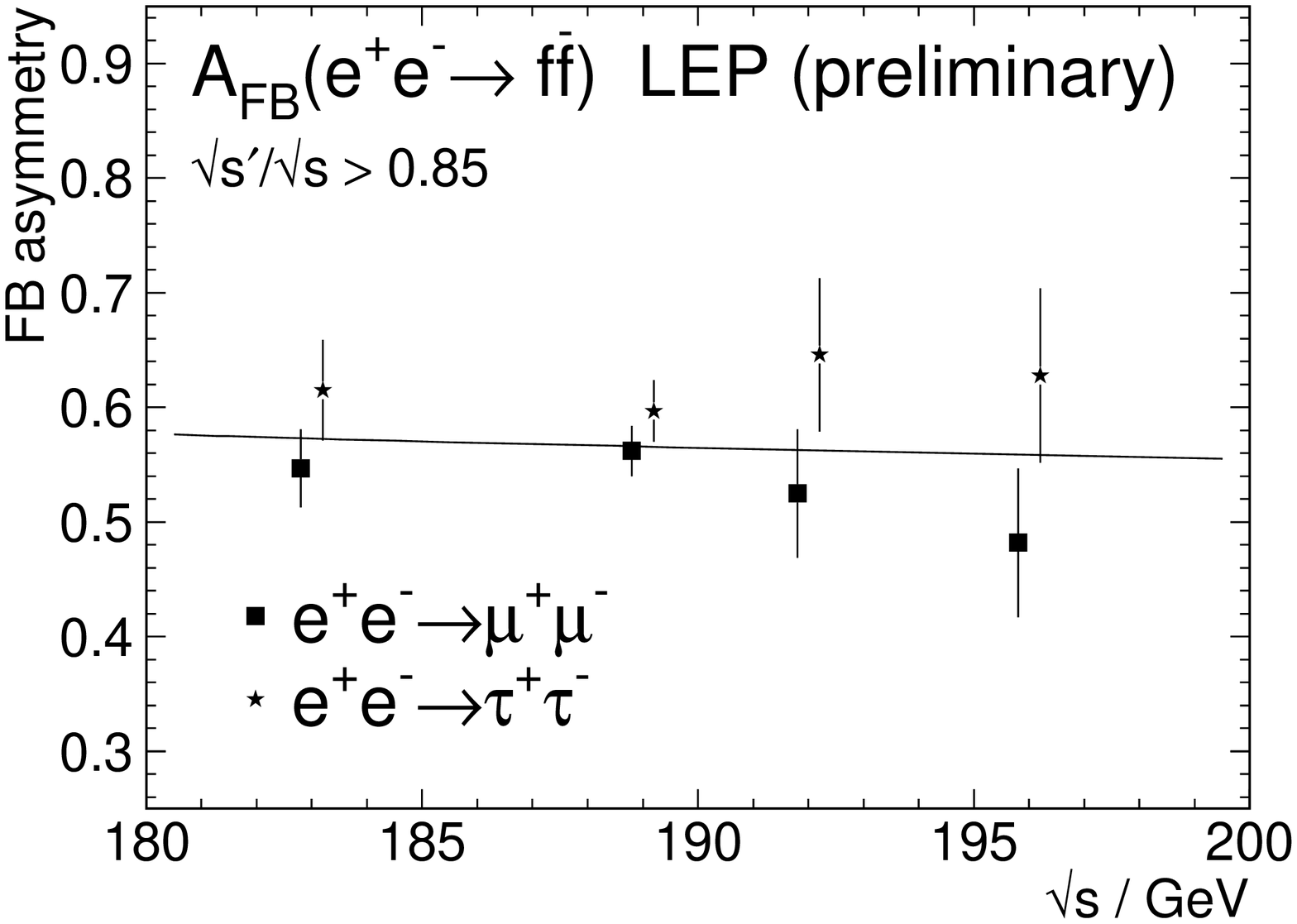,height=2.6in}
\caption{a) LEP average fermion-pair
cross-sections\,\protect\cite{ewwg} 
scaled by the Standard Model expectation, from ZFITTER\,\protect\cite{zfitter}; 
b) LEP average
forward-backward asymmetries, compared with the ZFITTER prediction.}
\label{fig:fflep}
\end{center}
\end{figure}

The cross-sections for fermion-pair production have been
measured\,\cite{alephtwofermion,delphitwofermion,l3twofermion,opaltwofermion} in
the hadronic channel (q$\overline{\mathrm{q}}$ production), and for
$\mu^+\mu^-$, $\tau^+\tau^-$ and $\ee$ (the latter dominated by
t-channel Bhabha scattering). 
The Standard Model expectation describes the data well,
as shown in Figure~\ref{fig:fflep}a for the combined LEP
cross-sections.

For muon and tau pair production, the easily identifiable lepton
charge is further employed to measure the forward-backward
asymmetry of the non-radiative events. 
As shown in Figure~\ref{fig:ffasy}, the asymmetry for non-radiative
events is large at these
energies, in contrast to that at LEP1. Again the Standard Model
expectation describes the data well -- an expanded view of the higher
energy LEP-averaged asymmetries is shown in Figure~\ref{fig:fflep}b.

In addition to the results presented, measurements have further been
made of heavy quark pair production at LEP2 
energies\,\cite{alephtwofermion,hfprod}. 
They too are well described by the Standard Model
expectation.
Constraints on a wide range of new physics scenarios have been placed
with the fermion-pair data, ranging from four-fermion contact
interactions to electroweak scale quantum gravity
\,\cite{vanina,alephtwofermion,delphitwofermion,opaltwofermion,newphys2f}. 
Discussion of these topics is beyond the scope of this report.

\section{QED Tests}

A few electroweak processes at LEP2 do not have any
significant contribution from massive vector boson exchange, and
so may be employed to test the adequacy of quantum electrodynamics,
QED, at the highest LEP energies.

Tests have been made with the process 
$\ee\to\gamma\gamma(\gamma)$\,\cite{photonics}.
Possible deviations from the QED
expectation are parameterised in terms of an effective cut-off
parameter $\Lambda_{\pm}$\,\cite{drell}. 
Typical limits obtained by each experiment are
$\Lambda_{\pm}\geq$~290~GeV at 95\% CL.

\section{W Pair Production and Decays}
\label{sec:sigww}

\begin{figure}[htb]
\begin{center}
\epsfig{file=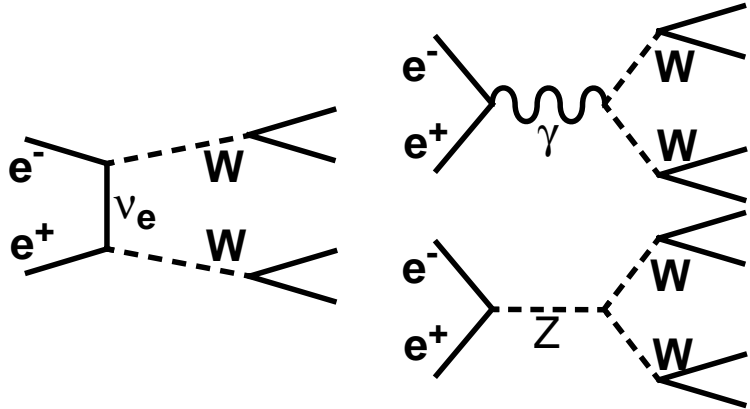,height=2in}
\caption{The three W pair production diagrams. The fermions from W
decay are also shown.}
\label{fig:cc03}
\end{center}
\end{figure}

\begin{figure}[thb]
\begin{center}
\epsfig{file=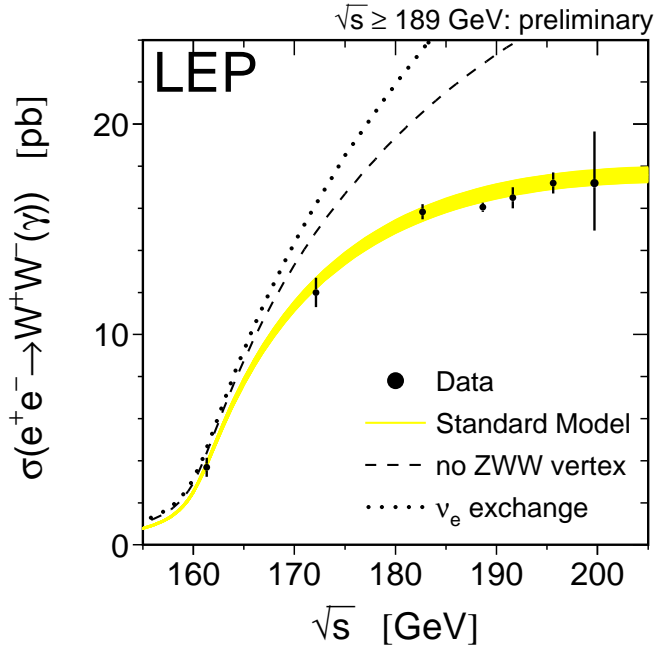,height=3.4in}
\caption{LEP average W pair production
cross-section\,\protect\cite{wwsel,ewwg}, corrected to
correspond to the three doubly-resonant W pair production diagrams 
(CC03). The predicted cross-sections are from
GENTLE\,\protect\cite{gentle}.}
\label{fig:sigww}
\end{center}
\end{figure}

At LEP2, three diagrams contribute to doubly-resonant W pair
production, as shown in Figure~\ref{fig:cc03}.
The neutrino exchange diagram dominates close to the W pair
threshold, and in the Standard Model the main effect of the other
two diagrams at LEP energies is a negative interference.
This is illustrated in Figure~\ref{fig:sigww}, where the expected
cross-section is shown with the full Standard Model structure, and if
one or both of the diagrams with triple vector boson couplings
is omitted. The effect of the triple gauge coupling is discussed
further in section~\ref{sec:tgc}.

\begin{table}[b]
\begin{center}
\begin{tabular}{l|cc}  
Decay mode &  Efficiency &  Purity \\
\hline
$\WW\to\qqqq$ & 90\% & 80\% \\
$\WW\to\qqln$ & 82\% & 90\% \\
$\WW\to\lnln$ & 60-80\% & 90\% \\
\end{tabular}
\caption{Typical W pair event selection efficiencies and purities.}
\label{tab:wwsel}
\end{center}
\end{table}

The typical selection efficiencies and purities for W pair
events\,\cite{wwsel} are
given in Table~\ref{tab:wwsel}. 
The main backgrounds arise from other four-fermion processes, 
and non-radiative $\qq$ events
in the $\qqqq$ and $\qqln$ decay channels, or lepton-pair
production and multiperipheral interactions in the
$\lnln$ channel. 
The cross-sections measured\,\cite{wwsel,ewwg} are shown in 
Figure~\ref{fig:sigww} for
the various LEP2 centre-of-mass energies. 
The peak cross-section, of around 17~pb, is more than three orders of
magnitude less than the Z cross-section at LEP1: consequently even with
the high luminosities collected at LEP2 the W pair events
number a few thousand per experiment.
Nonetheless, precision electroweak measurements can be made.
 
\begin{table}[b]
\begin{center}
\begin{tabular}{l|c}  
Decay mode & Branching ratio (\%) \\
\hline
W~$\to\mathrm{e}\nu$ & 10.61$\pm$0.25 \\
W~$\to\mu\nu$        & 10.65$\pm$0.24 \\
W~$\to\tau\nu$       & 10.82$\pm$0.32 \\
\hline
W~$\to\ell\nu$       & 10.68$\pm$0.13 \\
W~$\to\mathrm{q}\overline{\mathrm{q}}$    & 67.96$\pm$0.41 \\
\end{tabular}
\caption{LEP average W decay branching ratio measurements\,\protect\cite{wwsel,ewwg}. 
Results are preliminary.}
\label{tab:wwbr}
\end{center}
\end{table}

The branching ratio for W decays via the electron, muon, tau and
hadronic modes have been measured by all four experiments\,\cite{wwsel}. 
The LEP average results are given in Table~\ref{tab:wwbr}. 
The results for the individual leptonic channels are consistent with
lepton universality, and the average leptonic branching ratio is also
consistent with the Standard Model expectation.
The precision of the measurement of B(W~$\to\ell\nu$) from LEP is now
better than that from $\pp$ colliders.

The leptonic W branching ratio can be re-interpreted\,\cite{wwsel} 
in terms of the
CKM matrix element $\Vcs$ without need for a CKM unitarity
constraint, using the relatively well-known values of
other CKM matrix elements involving light quarks\,\cite{pdg98}.
These indirect constraints currently lead to a value\,\cite{ewwg} of
$|\Vcs|=0.997\pm0.020$, much more precise than the value derived from D
decays of 1.04\,$\pm$\,0.16\,\cite{pdg98}.

\section{Measurement of the W Mass and Width}

At centre-of-mass energies above the W pair threshold, the technique
for measuring the W mass lies in the reconstruction of the directions
and energies of the four primary W decay products.
These may be either four quarks, approximated by four jet 
directions and energies, for the $\qqqq$ channel;
or two quarks/jets and a charged lepton for the $\qqln$ channel,
deducing the neutrino direction and energy from the missing momentum
in the event.
Decays to $\lnln$ are of limited use because at least two neutrinos
are undetected.
The W decay products are paired up to give reconstructed W mass
estimates. 
A substantial improvement is made in the mass resolution for both
$\qqqq$ and $\qqln$ 
channels by applying a kinematic fit, constraining the total
energy and momentum in the event to be that of the known colliding
electron-positron system, making a small correction for possible
unobserved initial-state radiation.

\begin{figure}[htb]
\begin{center}
\epsfig{file=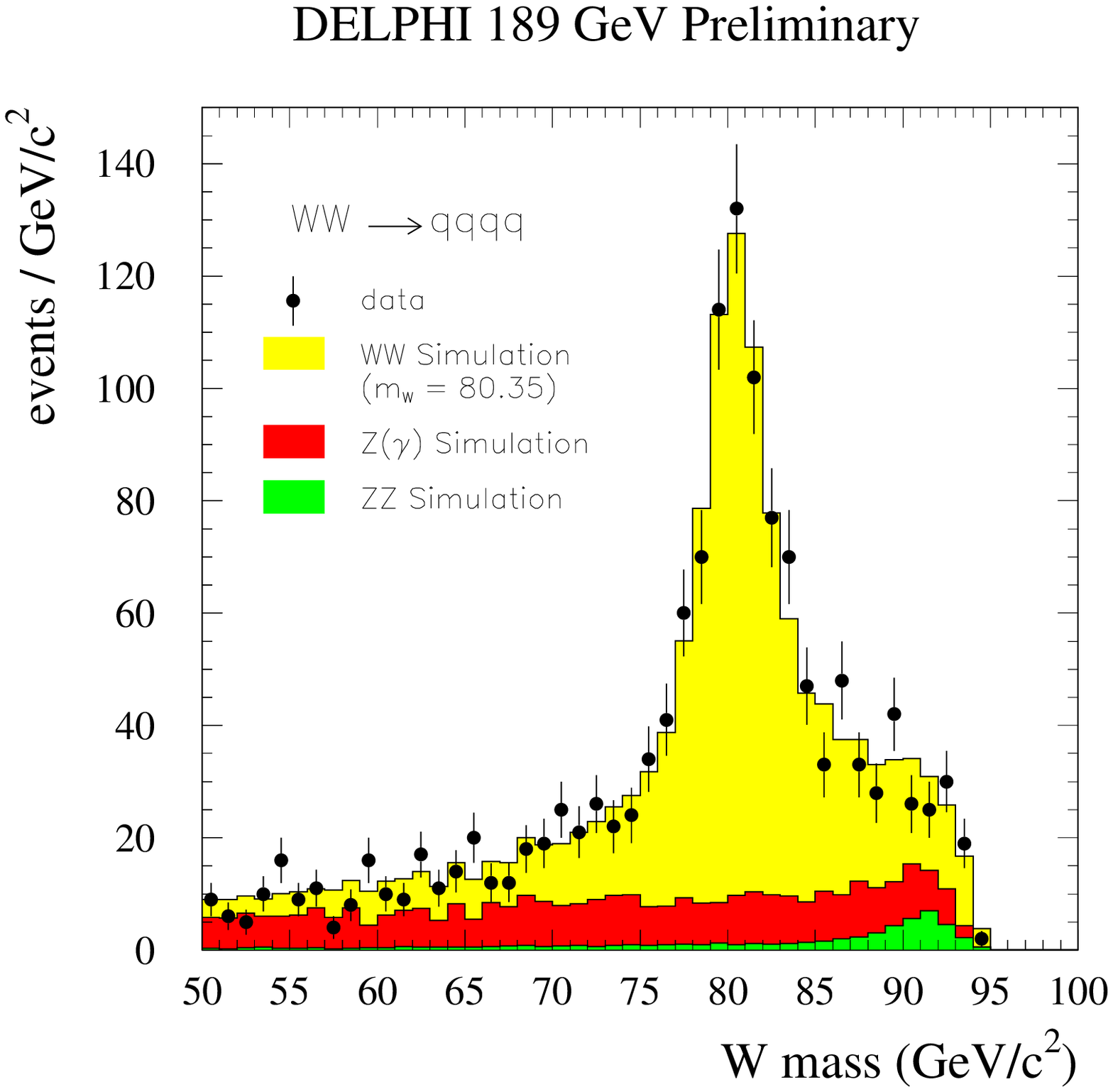,height=2.2in}
\epsfig{file=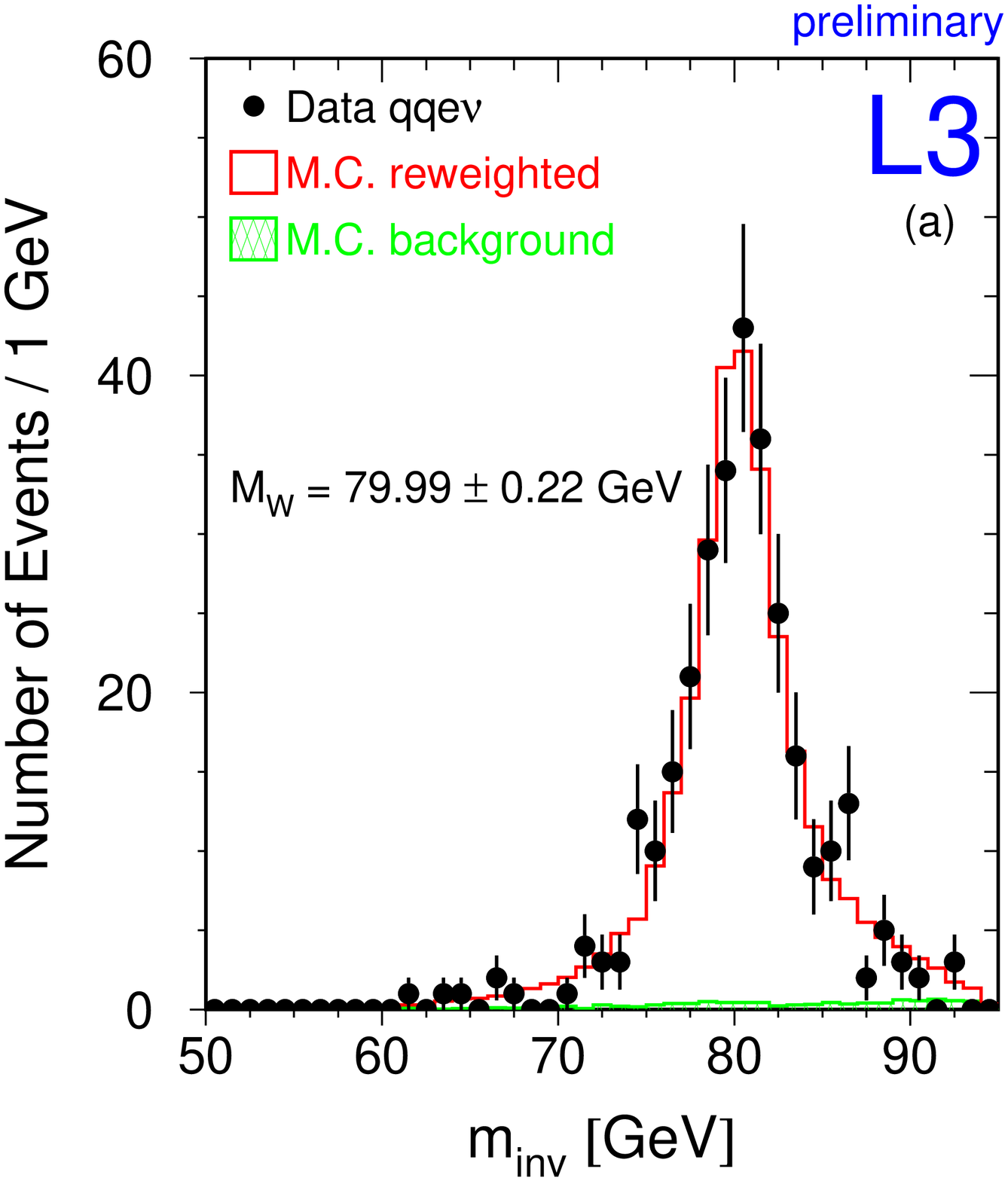,height=2.2in} \\
\epsfig{file=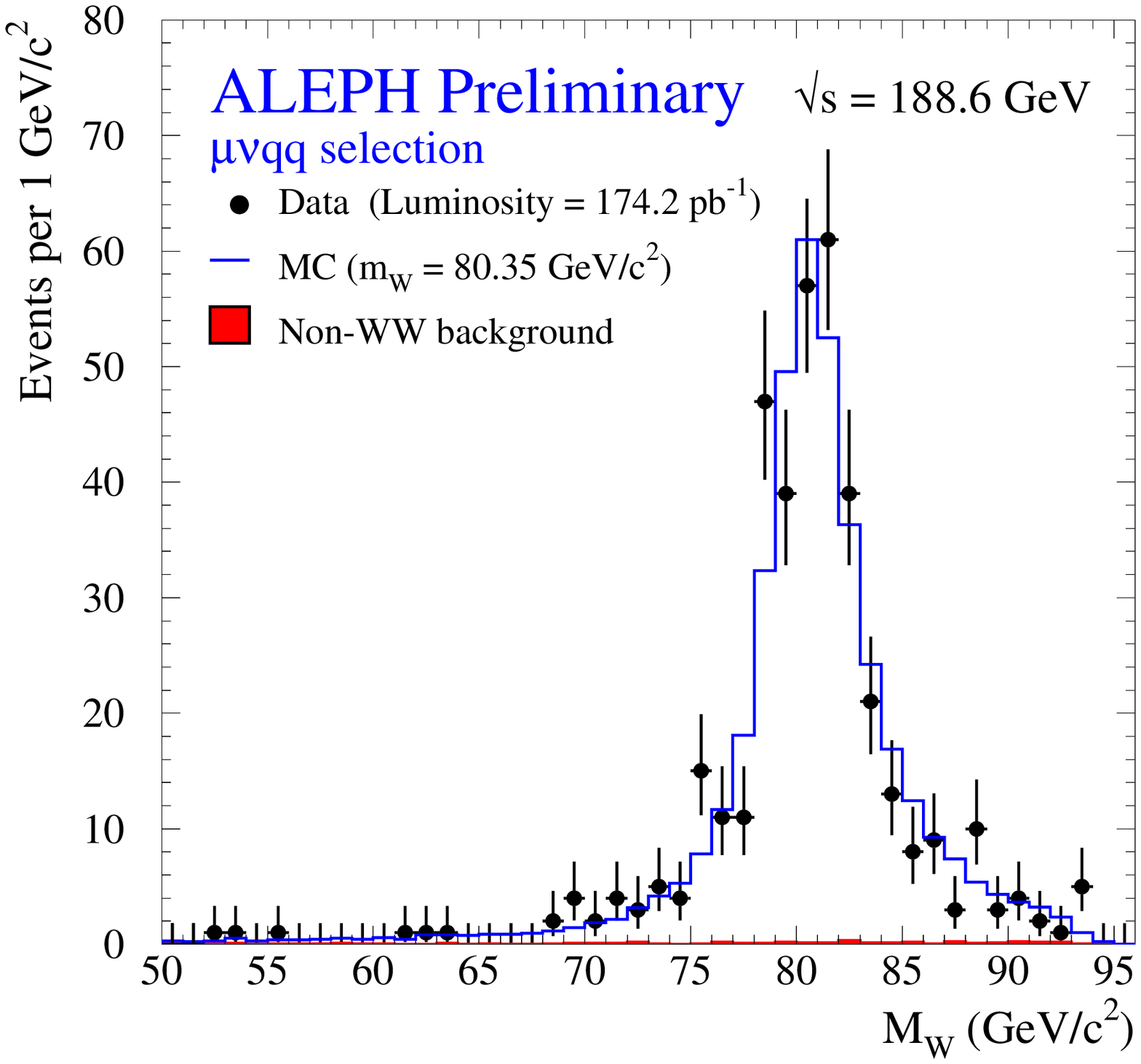,height=2.2in}
\epsfig{file=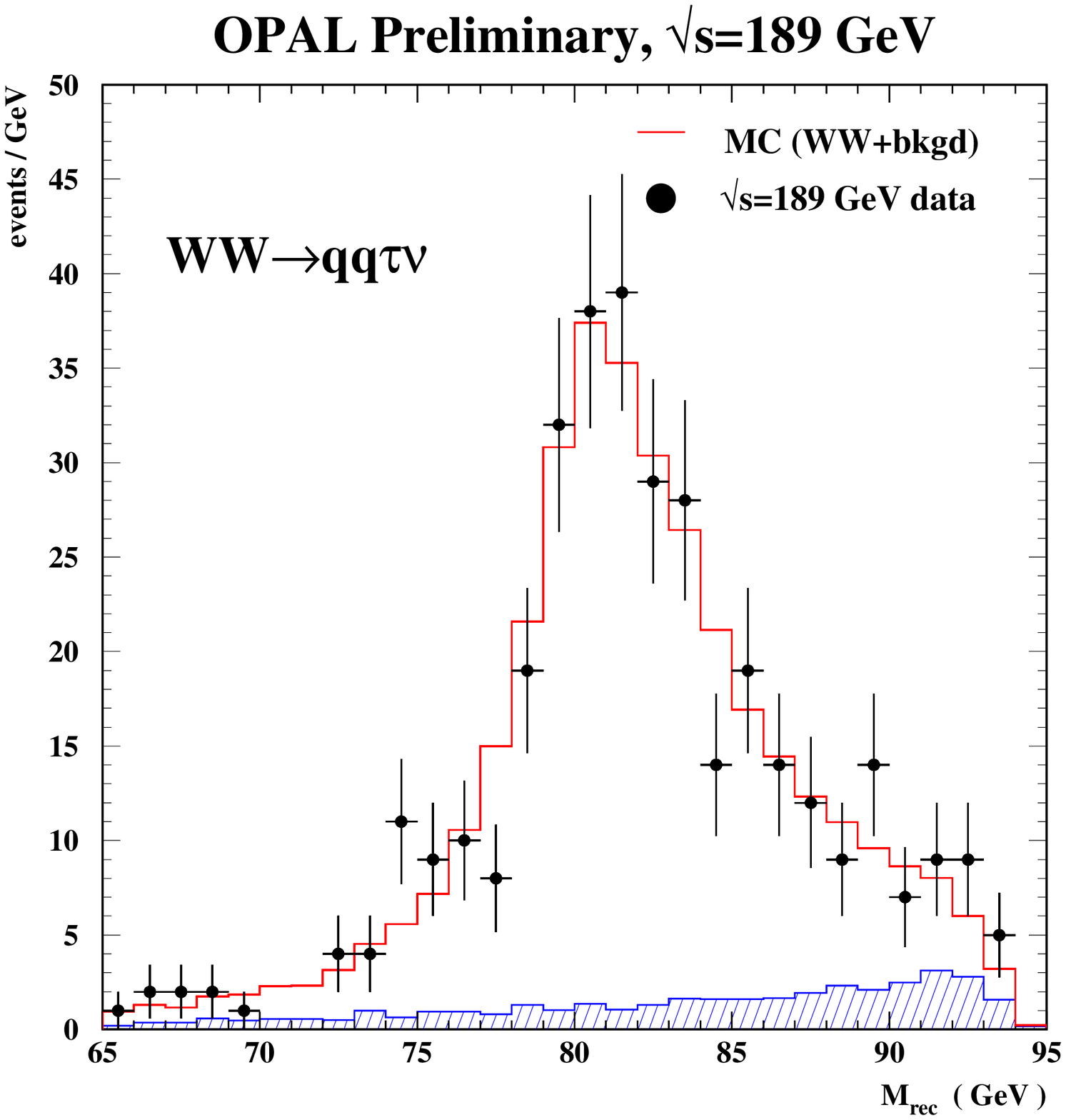,height=2.2in}
\caption{Reconstructed W mass distributions obtained from the
kinematic fits, for the four-quark and three $\qqln$ WW decay channels\,\protect\cite{mw,Wwidth}.}
\label{fig:mw}
\end{center}
\end{figure}
Typical reconstructed mass distributions from the kinematic fit are
shown in Figure~\ref{fig:mw}, for the $\qqqq$ and three $\qqln$
channels.
Clear W mass peaks are observed, in some cases with very low
backgrounds.
The W mass is extracted\,\cite{mw,Wwidth} 
from the measured W masses in each data event
using a Monte Carlo technique, the details of which differ from one
experiment to another. 
The Monte Carlo techniques have in common that they use full detector
simulations to correct for the effects of finite detector acceptance
and resolution, as well as initial-state radiation, and in most cases
a reweighting technique is used to model different true W mass values.
The results obtained from the fits are given in
Table~\ref{tab:wmass}, separately for the $\qqqq$ and $\qqln$ channels
(in the case of ALEPH, an analysis employing also purely leptonic W
decays is included).
\begin{table}[b]
\begin{center}
\begin{tabular}{l|cc}  
           & \multicolumn{2}{c}{Mass measurement (GeV)} \\
Experiment & $\WW\to\qqqq$ & $\WW\to\qqln$($\lnln$) \\ 
\hline
ALEPH  & 80.561$\pm$0.121 & 80.343$\pm$0.098 \\
DELPHI & 80.367$\pm$0.115 & 80.297$\pm$0.155 \\
L3     & 80.656$\pm$0.156 & 80.224$\pm$0.135 \\
OPAL   & 80.345$\pm$0.134 & 80.656$\pm$0.156 \\
\hline
LEP    & 80.429$\pm$0.089 & 80.313$\pm$0.063 \\
\hline
LEP (161 GeV)& \multicolumn{2}{c}{80.40$\pm$0.22} \\
Combined     & \multicolumn{2}{c}{80.350$\pm$0.056} \\
\end{tabular}
\caption{LEP W mass results from the different channels and
experiments\,\protect\cite{mw,Wwidth,ewwg}.
All results except those from 161 GeV are preliminary.}
\label{tab:wmass}
\end{center}
\end{table}
For all of these fits, the W width is taken to have its expected
Standard Model dependence on the W mass.
The results are consistent with each other, and also with the W mass
extracted from the measurement of the W pair threshold cross-section
at 161~GeV.
The overall LEP average W mass measurement obtained is 
thus\,\cite{ewwg}:
\beq
\mw = 80.350 \pm 0.056 \mbox{~GeV~~~(LEP)}
\eeqn
\clearpage
The LEP W mass measurement is slightly more precise than that from $\pp$
colliders, of $80.448 \pm 0.062$~GeV\,\cite{markl}.
The two measurements have similar precision but use very different
techniques, and so are essentially uncorrelated. 
A substantial improvement is obtained by
averaging them\,\cite{ewwg}:
\beq
\mw = 80.394 \pm 0.042 \mbox{~GeV~~~(World Average)}
\eeqn

The width of the W mass distributions shown in Figure~\ref{fig:mw} has
components from the true W width and from detector resolution. In
many events the mass resolution is comparable to, or better than, the
true width. 
It is consequently possible to measure directly both the
W mass and width, and in practice the two results are little
correlated.
This has been done by three experiments\,\cite{Wwidth}, the combined
result currently being:
\beq
\gw = 2.12 \pm 0.20 \mbox{~GeV.}
\eeqn
With the full LEP2 statistics the precision should improve on 
the current measurement from CDF, of
2.055$\pm$0.125~GeV\,\cite{markl}.

\begin{figure}[htb]
\begin{center}
\epsfig{file=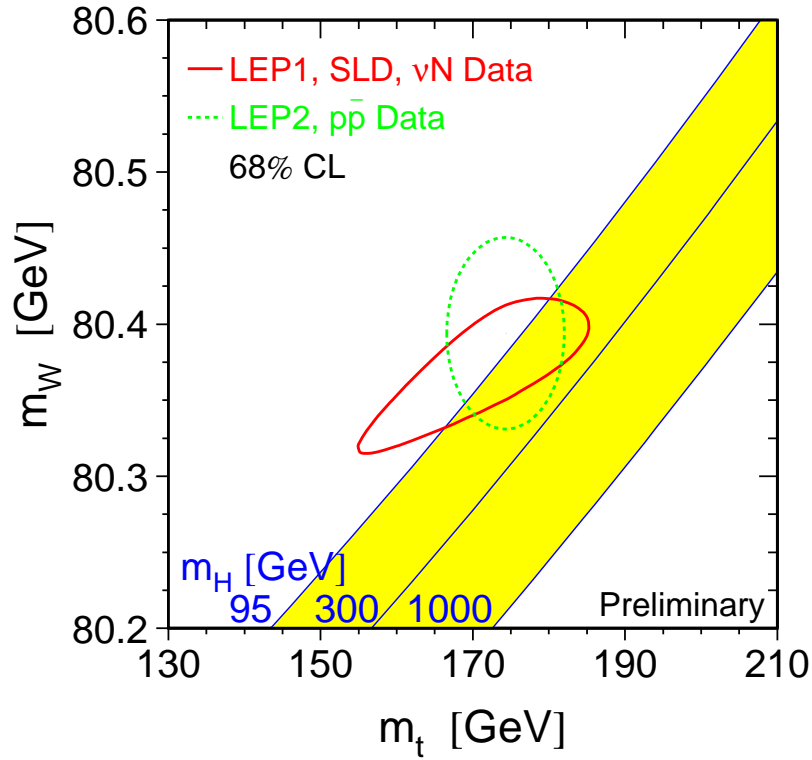,height=4in}
\caption{Dashed contour: direct measurements of W and top masses;
Solid contour: predicted values of $\mw$ and $\mt$ from LEP1, SLD and
neutrino-nucleon scattering precise electroweak data\,\protect\cite{ewwg}.
The diagonal band indicates the expected Standard Model 
dependence for different Higgs boson masses.}
\label{fig:mwmt}
\end{center}
\end{figure}

With the current uncertainties, the measurement of the W mass starts to
provide an interesting further test of the Standard Model relative to
other precision electroweak measurements.
This is illustrated in Figure~\ref{fig:mwmt}\,\cite{ewwg}: 
the predicted W and top masses extracted from fits to lower energy
electroweak data are consistent with the direct measurements from LEP
and the Tevatron\,\cite{markl}, 
and the precision of the measurements is
similar to that of the prediction.
From the overlaid curves showing the Standard Model expectation as a
function of the Higgs boson mass, it is further evident that both the
precise lower energy measurements, and the direct W and top mass 
measurements taken together, 
separately favour a Standard Model Higgs boson in the relatively low
mass region.

At the time of writing, the LEP W mass analyses are in a stage 
of detailed review and improvement, as careful systematic studies 
are needed to match the statistical precision: for this reason 
the W mass results of all experiments are preliminary.
It is not possible to predict with certainty the final W mass error 
from LEP2, but it is interesting to consider the main error sources, 
in order to extrapolate to the full data sample.
If the present LEP average W mass result is broken down into
statistical and systematic components, they are respectively
approximately 36~MeV and 43~MeV.
This does not mean simply that a systematic limit is being reached,
because the analysis is performed in two channels, $\qqqq$ and
$\qqln$, of approximately equal statistical weight but with different
systematic error behaviour.
In the $\qqln$ case the main systematic errors arise from detector 
calibration uncertainties and Monte Carlo statistics:
these are amenable to reduction with more statistics, and are
uncorrelated between the different LEP experiments. 
In neither case do they give a large contribution to the combined LEP
error.
In the $\qqqq$ channel, on the other hand, the systematic errors are
dominated by contributions from final-state interaction effects such
as colour-reconnection\,\cite{CR} and Bose-Einstein
correlations\,\cite{BEC}, and the
modelling of backgrounds and fragmentation uncertainties. 
These error sources are largely correlated between
different experiments, and will be relatively difficult to reduce.
Consequently, with the full LEP2 data sample 
the $\qqln$ channel analysis should remain statistics
limited, but the $\qqqq$ channel will probably be limited by
systematics at some level below the currently estimated error. 
A realistic expectation for the overall combined W mass error from
LEP2 then lies in the region of 30~MeV.

\section{Gauge Boson Self-Interactions}

\label{sec:tgc}

\begin{figure}[thb]
\begin{center}
\raisebox{2.4in}{a)}
\epsfig{file=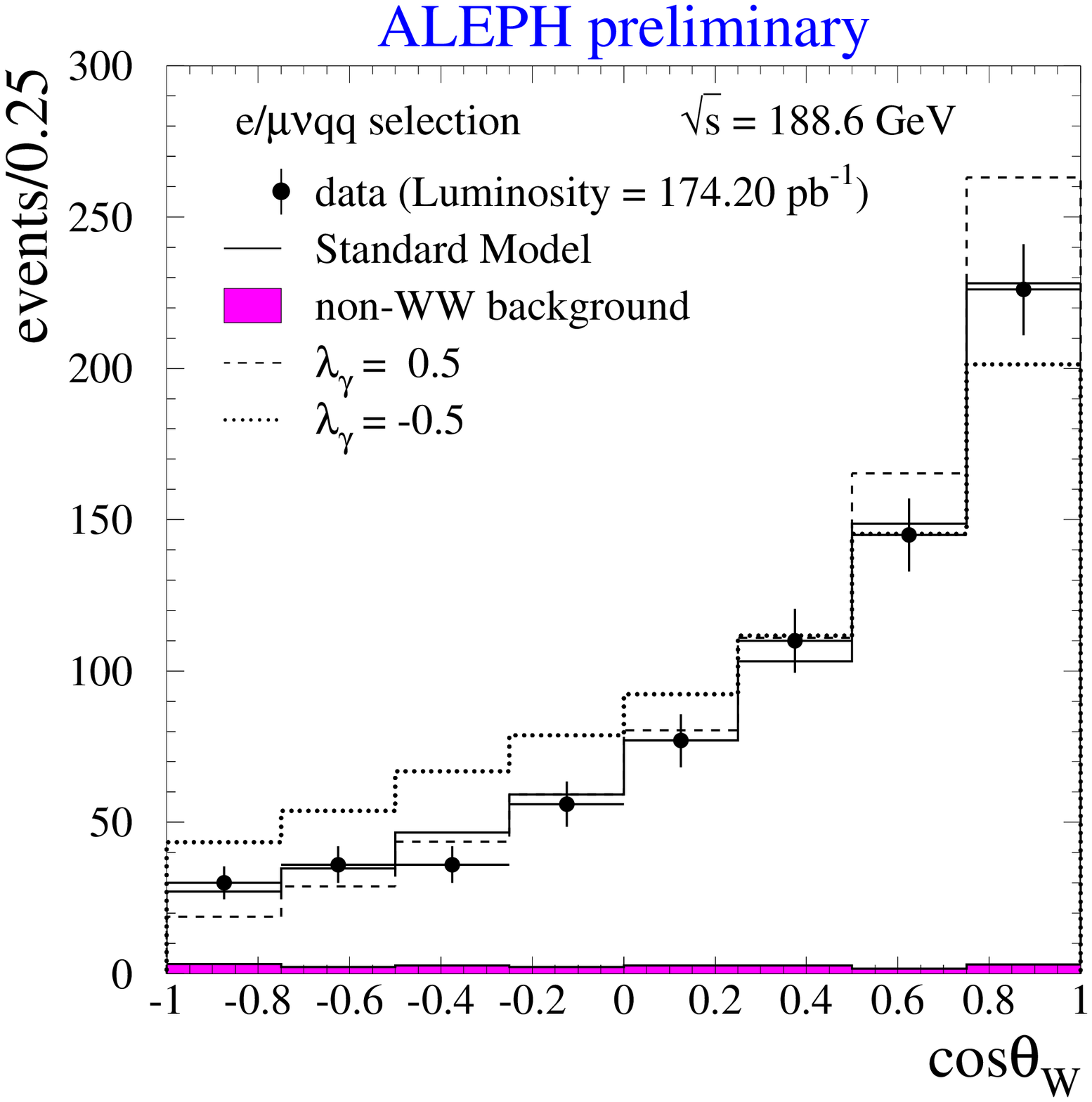,height=2.5in}
\raisebox{2.4in}{b)}
\epsfig{file=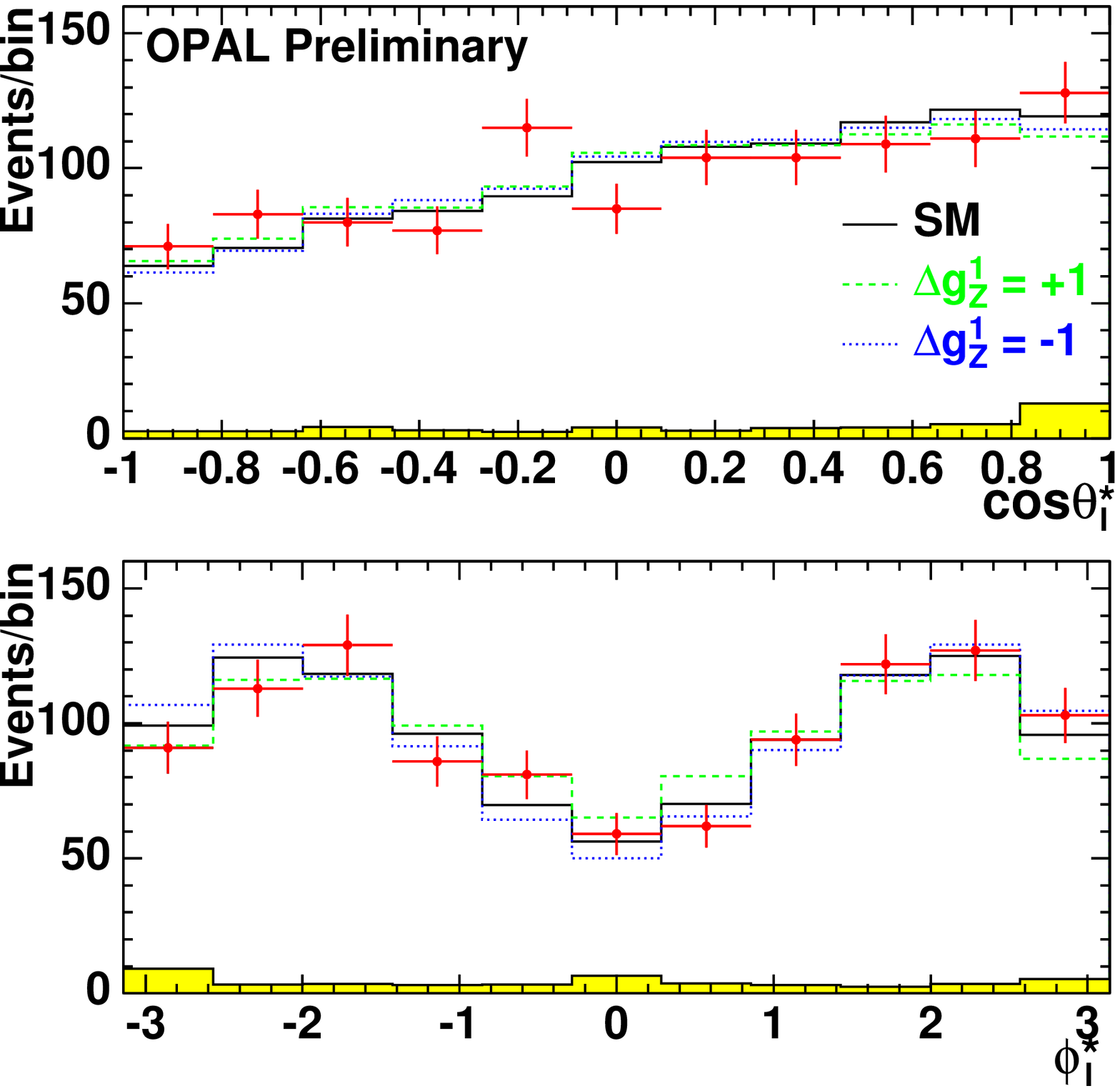,height=2.5in}
\caption{Reconstructed W (a) production\,\protect\cite{alephtgc} and (b) decay angle
distributions\,\protect\cite{opaltgc}, compared to the Standard Model expectation and
different anomalous coupling hypotheses.}
\label{fig:wang}
\end{center}
\end{figure}

As mentioned in Section~\ref{sec:sigww}, W pair production at LEP2
probes the triple gauge boson vertices WW$\gamma$ and WWZ.
The form and strength of these vertex couplings is unambiguously
predicted by the gauge structure of the Standard Model.
Possible new physics beyond the Standard Model may additionally 
bring in extra effective interactions between three gauge bosons without
affecting other sectors, mandating a test of this aspect
of the theory.

The general lagrangian for the WWV (V=$\gamma$ or Z) interaction
contains 14 parameters. 
Constraints of C, P and CP invariance and U(1)$_{em}$ gauge invariance
reduce these to five parameters, and constraints from low energy
measurements reduce these further to three, conventionally taken to be 
$\kag$, $\goz$ and $\lag$, respectively 1, 1 and 0 in the Standard
Model.
The LEP experimental analyses are performed in terms of these three
variables, or equivalently deviations from their respective Standard
Model values.

The triple gauge couplings affect the characteristics of W pair
production in three ways:
the total cross-section changes, as shown in
Figure~\ref{fig:sigww}, by an amount which increases rapidly with
$\sqrt{s}$; 
the production angular distribution of the W is modified, as
shown in Figure~\ref{fig:wang}a; 
finally, the helicity mixture of the Ws produced at a
given $\cos\theta$ is affected. 
This last effect can be measured experimentally by using the W decay
as a polarisation analyser. 
Typical W decay angle distributions under different coupling
hypotheses are shown in Figure~\ref{fig:wang}b.

Values of triple gauge coupling parameters are extracted from the W
pair data using the W production and decay 
angles\,\cite{alephtgc,delphitgc,l3tgc,opaltgc}.
This typically employs so-called optimal observables\,\cite{OO}
constructed from these angles, which has the advantage of
not requiring analysis of a five-dimensional differential distribution.
The preliminary results obtained, averaging over all experiments and
including
also the less sensitive single-W and $\nu\nu\gamma$
constraints\,\cite{delphitgc,othertgc}, 
are\,\cite{ewwg}:
\beqa
\kag & = & \phantom{-}1.04 \pm 0.08 \\
\goz & = & \phantom{-}0.99 \pm 0.03 \\
\lag & = & -0.04 \pm 0.04
\eeqan
strikingly well described by the Standard Model predictions of unity,
unity and zero, respectively.
The results are quoted for the case where only one of the
anomalous parameters differs from the Standard Model
at a time: fits have also been performed allowing
up to three parameters to vary at once: consistent results are
obtained.

An alternative perspective on the W-pair production process is
provided by analyses which directly measure the relative rates of
production of transversely and longitudinally polarised W bosons.
A recent analysis along these lines by L3\,\cite{l3wpol} gives the
fraction of longitudinal W polarisation as (24.4$\pm$4.8$\pm$3)\% at
189 GeV. Overall, longitudinal W polarisation is established at the
five standard deviation level. This contrasts with W production from
the $\qq\to W$ process at $\pp$ colliders, where the W is 
transversely polarised.

Recently a study has been carried out by OPAL\,\cite{qgc} of
the quartic gauge couplings between WW$\gamma\gamma$ and
WWZ$\gamma$. These couplings are non-zero in the Standard Model, but
the effect on the data is tiny for the LEP2 sample.
However, constraints have been placed on possible large anomalous
values of these parameters using WW$\gamma$ production with
photon energies above 10~GeV, and also the $\nu\overline{\nu}\gamma\gamma$
final-state, where there is sensitivity from the W fusion diagrams.
This analysis places the first, albeit weak, direct limits on quartic
gauge couplings.

\section{Z Pair Production}

\begin{figure}[htb]
\begin{center}
\epsfig{file=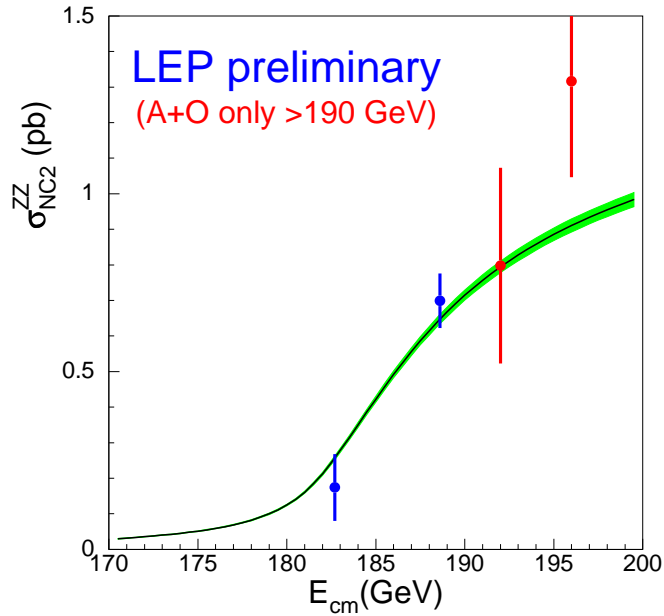,height=3.2in}
\caption{Observed Z pair production cross-section as a function of
centre-of-mass energy, compared to the Standard Model 
prediction from YFSZZ\,\protect\cite{yfszz}.}
\label{fig:sigzz}
\end{center}
\end{figure}

Since 1997, LEP has been running at centre-of-mass energies around and
above the Z pair production threshold. 
Unlike the W case, Z pair production involves no triple gauge coupling
diagrams in the Standard Model, but instead simply those with $t$ and
$u$ channel electron exchange.
Production is suppressed by factors of $(1-4\sin^2\theta_W)$
at two eeZ vertices, so that the Z pair cross-section is significantly
lower than that for W pairs.
The measured cross-section\,\cite{sigzz,ewwg} is shown in
Figure~\ref{fig:sigzz},
compared to the Standard Model prediction.
A particular interest of this process is that it forms an irreducible
background to potential Higgs boson production if the Higgs mass were
around the Z mass, in cases where one Z decays to b quarks. 
Figure~\ref{fig:sigzz} indicates that ZZ production is well understood.

\section{Summary}

With the excellent performance of the LEP machine at high energy in
the last couple of years, electroweak physics at LEP2 now truly
merits the epithet ``precise''.
The core measurements of the LEP2 programme, the W mass and the vector
boson self-couplings, have been made with precision better,
in some cases substantially so, than elsewhere.
Tests of the Standard Model with other processes serve to confirm the
superb description it provides of the data.

Finalisation of the current analyses, and inclusion of the 1999 and
2000 data samples, will provide significant further improvements in
precision, although requiring care and attention to the encroaching
systematic difficulties.
By the time of the next Lepton-Photon meeting, the full fruits of this
labour should be harvested.

\bigskip
Much credit is due to the LEP electroweak working group for the
preparation of most of the averages and figures presented. The
work of this team makes the job of a rapporteur simpler
and more enjoyable.
For their help during the preparation of this talk,
I wish to thank:
R.\,Bailey,
R.\,Clare,
J.\,Ellison,
M.\,Gr\"unewald,
R.\,Hemingway,
M.\,Hildreth,
E.\,Lan\c{c}on,
M.\,Lancaster,
C.\,Matteuzzi,
D.\,Miller,
K.\,M\"onig,
D.\,Plane,
A.\,Straessner,
D.\,Strom,
M.\,Thomson,
H.\,Voss,
P.\,Ward and
P.\,Wells.

\def\Discussion{
\setlength{\parskip}{0.3cm}\setlength{\parindent}{0.0cm}
     \bigskip\bigskip      {\Large {\bf Discussion}} \bigskip}
\def\speaker#1{{\bf #1:}\ }

\Discussion

 \speaker{Howie Haber (UCSC)}
Could you comment on the maximum energy achievable at LEP?

\speaker{Charlton} Increasing the energy beyond 200 GeV will be difficult, 
and it is 
unclear
how the performance will evolve.  An absolute maximum is 205-206 GeV, 
but 202-203 GeV
may be more realistic.

\speaker{Tom Ferbel (University of Rochester)}
You mentioned the observation of longitudinal W polarization at LEP2. 
There is also dominant
longitudinal W production in top quark decay, and this has been 
reported by the Tevatron experiments.

\speaker{Charlton}  Yes, that is correct. 
The significance observed in that case, however, is at
a much lower level than that from LEP.

\speaker{Michael Peskin (SLAC)}
You quoted a large systematic error for
the W mass measurement in
WW$\to\qqqq$---50-90 MeV---due to  color 
reconnection and
Bose-Einstein
effects.  Models of color reconnection predict observable
manifestations in the final state, and since these are not observed, 
the constraints should put bounds on these 
errors.  Could you comment on this?

\speaker{Charlton}
Some models of colour-reconnection effects have been excluded by LEP
data; however, others will be hard to test even with the full LEP2 
statistics.  The
errors I quoted also have large Monte Carlo statistical components.  A  final
error on the $\qqqq$ channel from this source of around 30 MeV may be
achievable.

\end{document}